\documentclass{PoS}

\newcommand{\rb}[1]{\raisebox{1.5ex}[-1.5ex]{#1}}
\newcommand{\swift}{{\it Swift}}

\newcommand{\xmm}{{\it XMM-Newton}}

\title{Finding AGN in Deep X-ray Flux States with Swift}

\ShortTitle{Finding AGN in Deep X-ray Flux States with Swift}

\author{Dirk Grupe   \\ %\thanks{A footnote may follow.}\\MPE\\
Department of Earth and Space Science, Morehead State University, 235 Martindale
Dr., Morehead, KY 40351, USA
E-mail: \email{d.grupe@moreheadstate.edu}}

\author{\speaker{S. Komossa} \\
 Max-Planck-Institut f\"ur Radioastronomie, Auf dem Hügel 69, 
 53121 Bonn, Germany\\
 E-mail: \email{skomossa@mpifr-bonn.mpg.de}}

\author{Mason Bush, Chelsea Pruett, Sonny Ernst, Taylor Barber, Jen Carter\\
Department of Earth and Space Science, Morehead State University, 235 Martindale
Dr., Morehead, KY 40351, USA; Rowan County Senior High School, Morehead, KY,
40351, KY, USA\\}

\author{Norbert Schartel,  Pedro Rodriguez, \& Maria Santos-Lleó \\
XMM-Newton Science Operations Centre, ESA, Villafranca del Castillo, Apartado 78, 
28691 Villanueva de la Cañada, Spain
}

\abstract{We report on	our ongoing project of finding Active Galactic Nuclei 
(AGN) that go into deep X-ray flux states detected by \swift. \swift\ is 
performing an extensive study on the flux and spectral variability of 
AGN using Guest Investigator 
and team fill-in programs followed by triggering \xmm\ for deeper 
follow-up observations. So far this program has been very successful and has 
led to a number of \xmm\ follow up observations, including  Mkn 335,
 PG 0844+349, and RX J2340$-$5329. Recent analysis of new \swift\ AGN 
 observations reveal several AGN went into a very low X-ray flux state,
 particularly Narrow-Line Seyfert 1 galaxies. One of these is RX J2317-4422, 
 which dropped by a factor of about 60 when compared to the ROSAT All-Sky
 Survey. 
}

\FullConference{Swift: 10 Years of Discovery\\
2-5 December 2014\\
La Sapienza University, Rome, Italy}

\begin{document}

\section{Introduction}
The \swift\ mission (Gehrels et al., 2004)
has revolutionized the ability to monitor and study Active
Galactic Nuclei (AGN). With its flexible scheduling and  multi-wavelength
capacity, \swift\ has enabled observers
to quickly react to changes discovered in the
fluxes of AGN. AGN often exhibit rapid and dramatic flux and spectral changes in
X-rays. These changes can be associated with absorption, reflection, or dramatic
changes in the accretion rate. Some of the most dramatic X-ray flux and spectral
changes have been observed in Narrow Line Seyfert 1 galaxies (NLS1s). This class
of AGN is characterized by relatively small black hole masses and high
accretion rates relative to the Eddington accretion rate. The most extreme of
these AGN is the NLS1 WPVS 007 which was bright in the X-ray sky during
the ROSAT All-Sky Survey (RASS)
in 1990. However, when observed a few years later with
ROSAT its X-ray flux had dropped by a factor of 400 (Grupe et al., 1995). 
During the RASS it WPVS 007  displayed the softest
X-ray spectrum of all bright X-ray detected AGN. 
FUSE observation in 2003
revealed strong absorption troughs in the UV spectrum of WPVS 007, which are
unseen in low-luminosity AGN. These deep absorption troughs are seen in Broad
Absorption line Quasars (BAL QSOs) and are the results of strong outflows.
WPVS 007 represents a link between BAL QSOs and NLS1s.

This research compares AGN X-ray fluxes found in the ROSAT All-Sky Survey
 with measurements from recent \swift\ observations. A decrease in the X-ray 
 flux of an AGN can result from absorption by gas, reflection on the 
 accretion disk, or a change in the accretion rate. 
Out of 80 AGN analyzed, six were found in a low X-ray flux state. 
A 1 ks \swift\ observation of each AGN confirmed our findings. One object, 
RX J2317$-$4422, was found to have decreased in X-ray flux by a factor of 
about 60 (Grupe et al., 2014). A follow up observation of RX J2317.8$-$4422 
with \xmm\ reveals an usually soft X-ray spectrum. 	

\section{Sample Selection and Swift}
 The Active Galactic Nuclei are selected from the ROSAT All-Sky Survey 
 (RASS) Bright Source Catalogue by Schwope et al. (2000). The total sample 
 size is  400 AGN, many of which have already been observed by \swift\ 
 multiple times (Grupe et al., 2010). Between 2014 April and September \swift\ 
 observed 80 of these AGN. 
 
The Space Science Center at Morehead State University in Eastern Kentucky has
an apprenticeship program that allows local senior
high school students to participate
in active research projects. In Fall 2014 
four high school students from the Rowan
County Senior High School participated in this program.
Each  student was given 20 AGN to check for significant 
  X-ray flux variability.

\section{Description of the Data Analysis}

To compare the \swift\ XRT count rates with those obtained during 
the RASS, dividing the RASS count rate by 4 is 
an appropriate estimate of the expected count rate. 
Students compared the expected \swift\ XRT count rate with the 
 measured \swift\ XRT count rate.  The XRT count rates were obtained by 
 using the \swift\ XRT analysis web-interface at the University of Leicester 
 \swift\ website
  (Evans et al., 2007, 2009; http://www.swift.ac.uk/user\_objects/;). 
  Students were task to find AGN which exhibited a change in 
  their count rate by factors of 5 or greater. Out of the sample of 80 AGN, 
  6 displayed a count rate that was at least a factor of 5 below the expected 
  value. The 6 AGN that exhibited shuch a large change in their count rate
  are listed in Table\,\ref{tab1}. 
  To confirm the X-ray low states, short 1 ks Swift ToO 
  observations were requested. Chelsea Pruett became the first high school 
  student ever requesting and obtaining a ToO observation with \swift. 
  All ToO observations confirmed the X-ray low state. 
  One AGN turned out to be very interesting: 
  RX J2317.8$-$4422, which displayed a decrease in its X-ray flux by a factor of
   20.

\begin{table} 
\begin{tabular}{lccc}
\hline
 & Expected \swift & Actual \swift & Factor \\
\rb{Name} & Count Rate & Count Rate & Change \\
\hline
RX J2317.8--4422 & 0.1775 & 0.010 & 17.8 \\
RX J2216.8--4451 & 0.3250 & 0.0325 & 10 \\
PKS 2310--322 & 0.085 & 0.010 & 8.5 \\
RX J1007.1+2203 & 0.155 & 0.025 & 6.2 \\
RX J1117.1+6522 & 0.1375 & 0.025 & 5.5 \\
MS 0117--28 & 0.135 & 0.025 & 5.4 \\
\hline
\end{tabular} 
\caption{List of AGN  which have been found to show a  decrease in their 
X-ray flux obtained by \swift\ compared with the ROSAT All-Sky Survey (RASS). 
The 
first column lists the source 
names, the  second the expected \swift\ XRT count rate estimated from the RASS 
count rate, the actually measured XRT count rate, and the last column lists 
the factors between the estimated and measured count rates.
} 
\label{tab1} 
\end{table}

\begin{figure}[h] 
\includegraphics[angle=270, width=.8\textwidth]{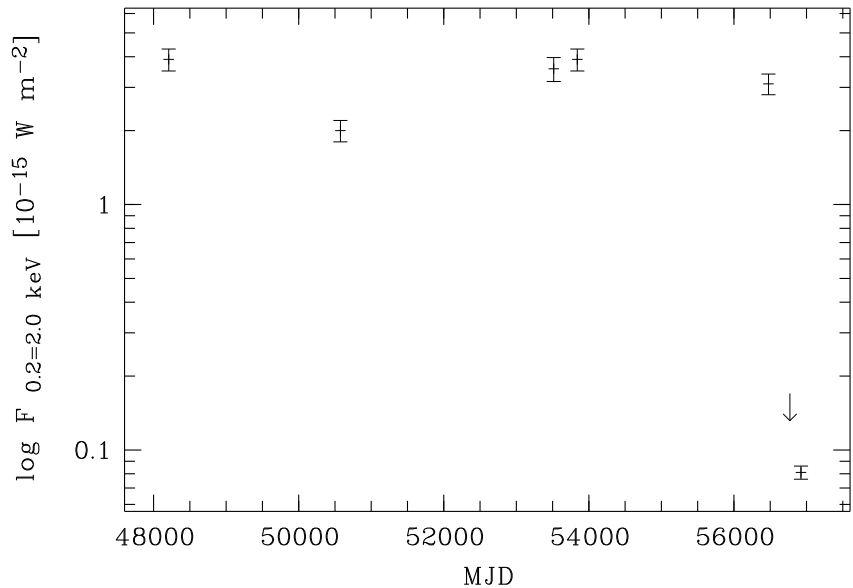} 
\caption{Long-term 0.2 - 2.0 keV light curve of RX J2317.8$-$4422.
 The first data point is the X-ray flux during the RASS and the second
  data point was derived from a ROSAT HRI observation (see Grupe et al., 2001).
   All other data points were derived from \swift\ observations starting in 2007 
   (Grupe et al., 2010).
} 
\label{rxj2317_long_lc} 
\end{figure}

\section{The NLS1 RX J2317.8-4422}

RX J2317.8$-$4422 (RA-2000: 23 17 49.9; Dec-2000: -44 22 28; z=0.132) is 
a Narrow-Line Seyfert 1 galaxy that has exhibited X-ray variability 
in the past with factors up to 7 (Grupe et al., 2001, 2010). 
 \swift\ observed this NLS1 in April 2014 and obtained a
3$\sigma$ upper limit of 1.76$\times 10^{-16}$ W m$^{-2}$, which means a drop by
a factor of more than 20 compared with the previous observation in August 2013.
Figure\,\ref{rxj2317_long_lc} displays the long-term 0.2-2.0 keV 
light curve of RX J2317.8$-$4422.  We triggered a \swift\ ToO observation in
September 2014 to confirm this X-ray flux low state and detected the AGN at a
level of 8$\times 10^{-17}$ W m$^{-2}$.

\begin{figure}[h] 
\includegraphics[angle=270, width=.8\textwidth]{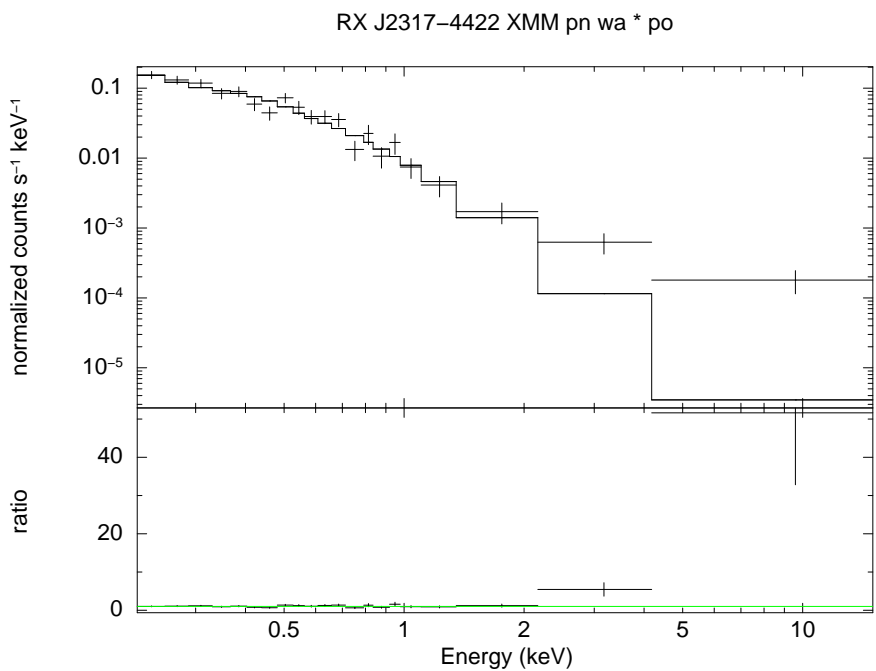} 
\caption{0.2-12 keV X-ray spectrum of RX J2317.8$-$4422 obtained by the  
EPIC pn onboard \xmm\ on 2014-October-29 fitted with a single power 
law model with $\alpha_{\rm x}$=2.81.} 
\label{rxj2317_xray_spec} 
\end{figure}

\begin{figure}[h] 
\includegraphics[angle=270, width=.95\textwidth]{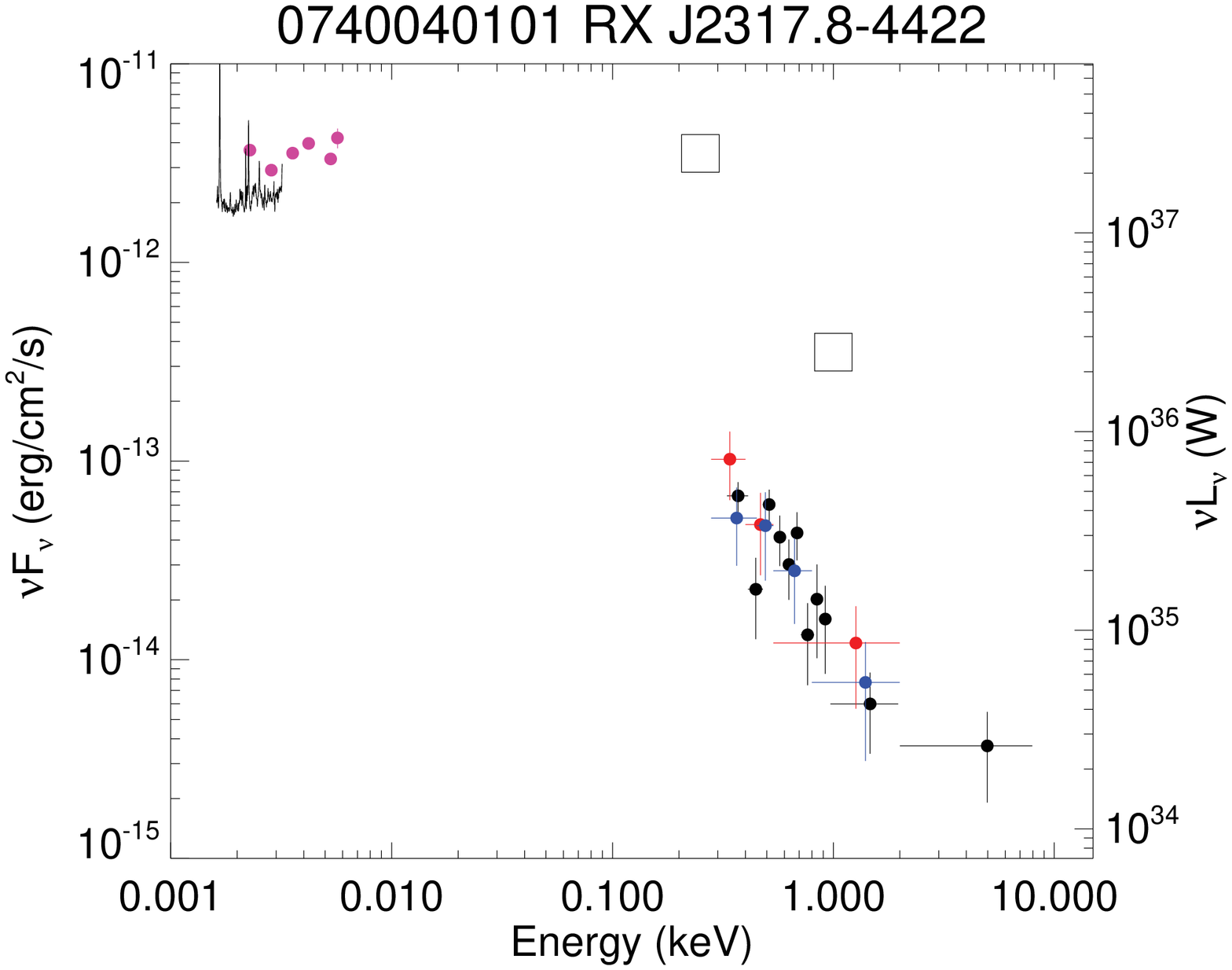} 
\caption{Spectral Energy Distribution of RX J2317$-$4422. The large squares 
display the X-ray data during the RASS. The optical/UV data points were 
derived from the \xmm\ OM, and the X-ray data were obtained from the EPIC pn 
and MOS detectors. The optical spectrum was taken in 1995 with the ESO 1.52m
telescope in La Silla/Chile
(Grupe et al., 1999, 2004). 
} 
\label{rxj2317_sed} 
\end{figure}

The extreme  X-ray flux change of RX J2317.8$-$4422
made it a valid target for an \xmm\ 
follow up program of AGN in deep minimum X-ray flux states. 
We obtained an observation with \xmm\ for 13 ks on 2014-October 29.
The X-ray spectrum obtained with the EPIC pn shown in
Figure\,\ref{rxj2317_xray_spec}
shows a very soft spectrum that can be modeled by a single power law 
model with an X-ray spectral slope of $\alpha_{\rm x}$=2.81$\pm$0.21. 
There is a hard X-ray component in the spectrum. 
To better understand the cause of this X-ray spectrum 
 a second \xmm\ observation was performed. A 100ks observation was performed by 
\xmm\ on 2014-November-17 simultaneously with \swift\ and NuStar. 
The second \xmm\ observation showed that the X-ray flux of 
 RX J2317.8$-$4422  had decayed even more and was found at a level of
 $(6.0\pm0.6)\times 10^{-17}$ W m$^{-2}$. 
Figure\,\ref{rxj2317_sed} compares the spectral 
energy distributions of RX J2317.8-4422 during the RASS (large squares)
 and the \xmm\ observation on October 29.

\section{Conclusions} 
Results of finding AGN in deep minimum states have again shown  that 
\swift\ has become a game changer in finding these unusual AGN. 
\swift\ is not only able to survey large samples of AGN very efficiently, 
it also allows fast follow up observations to verify the state of an AGN. 
Our \swift\ fill-in program of observing AGN  is ongoing and
 is expected to find more AGN with high X-ray variability in the future.
 The goal of the fillin program is to understand the changes in the spectral 
 energy distributions of AGN. We can conclude that these AGN in deep minimum 
 states  under extreme conditions.
 Our current understanding of the dramatic changes in the X-ray 
 flux and spectra is that this can be caused by absorption or X-ray 
 reflection (e.g. Grupe et al. 2012, Gallo et al. 2015). 
 A preliminary analysis of the \xmm\ observation of RX J2317.8--4422
 from November 2014 displays very soft X-ray spectrum, similar to that of WPVS
 007 during the RASS. Is RX J2317.8$-$4422 maybe the next link between NLS1s and
 BAL QSOs?
 We are currently
 working on the publication of the \swift, NuStar,
  and \xmm\ observations of RX J2317.8$-$4422.

\acknowledgments
This work made use of data supplied by the UK \swift\ Science Data Centre 
at the University of Leicester.  The work is based on observations obtained with
\xmm, an ESA science mission with instruments and contributions directly
funded by ESA Member States and NASA. We also want to thank Neil gehrels and the
\swift\ Mission Operation Center to make every effort to get our ToO requests
scheduled.

\end{document}